\documentclass[12pt,preprint]{aastex}

\begin{document}
\bibliographystyle{apj}

\newcommand{\lya}{Lyman-$\alpha$}
\newcommand{\eqw}{\hbox{EW}}
\newcommand{\zs}{z$\sim$6.96}
\newcommand{\La}{Ly$\alpha$}
\def\erg{\hbox{erg}}
\def\cm{\hbox{cm}}
\def\sec{\hbox{s}}
\def\f17{f_{17}}
\def\Mpc{\hbox{Mpc}}
\def\Gpc{\hbox{Gpc}}
\def\nm{\hbox{nm}}
\def\km{\hbox{km}}
\def\kms{\hbox{km s$^{-1}$}}
\def\year{\hbox{yr}}
\def\Myr{\hbox{Myr}}
\def\Gyr{\hbox{Gyr}}
\def\deg{\hbox{deg}}
\def\arcsec{\hbox{arcsec}}
\def\microJy{\mu\hbox{Jy}}
\def\zre{z_r}
\def\fesc{f_{\rm esc}}

\def\ergcm2s{\ifmmode {\rm\,erg\,cm^{-2}\,s^{-1}}\else
                ${\rm\,ergs\,cm^{-2}\,s^{-1}}$\fi}
\def\flam_unit{\ifmmode {\rm\,erg\,cm^{-2}\,s^{-1}\,\AA^{-1}}\else
                ${\rm\,ergs\,cm^{-2}\,s^{-1}\,\AA^{-1}}$\fi}
\def\fnu_unit{\ifmmode {\rm\,erg\,cm^{-2}\,s^{-1}\,Hz^{-1}}\else
                ${\rm\,ergs\,cm^{-2}\,s^{-1}\,Hz^{-1}}$\fi}
\def\ergsec{\ifmmode {\rm\,erg\,s^{-1}}\else
                ${\rm\,ergs\,s^{-1}}$\fi}
\def\kmsMpc{\ifmmode {\rm\,km\,s^{-1}\,Mpc^{-1}}\else
                ${\rm\,km\,s^{-1}\,Mpc^{-1}}$\fi}
\def\cMpc{\ifmmode {cMpc}\else
                ${cMpc}$\fi}
\def\kpc{{\rm kpc}}
\def\nv{\ion{N}{5} $\lambda$1240}
\def\civ{\ion{C}{4} $\lambda$1549}
\def\oii{[\ion{O}{2}] $\lambda$3727}
\def\oiipair{[\ion{O}{2}] $\lambda \lambda$3726,3729}
\def\oiii{[\ion{O}{3}] $\lambda$5007}
\def\oiiib{[\ion{O}{3}] $\lambda$4959}
\def\oiiipair{[\ion{O}{3}] $\lambda \lambda$4959,5007}
\def\taulya{\tau_{Ly\alpha}}
\def\taubar{\bar{\tau}_{Ly\alpha}}
\def\llya{L_{Ly\alpha}}
\def\ldlya{{\cal L}_{Ly\alpha}}
\def\nbar{\bar{n}}
\def\Msun{M_\odot} 
\def\mdyn{M_{dyn}}
\def\vmax{v_{max}}
\def\sqamin{\Box'}
\def\l43{L_{43}}
\def\ls{{\cal L}_{sym}}
\def\snrat{\ifmmode {\cal S / N}\else
                   ${\cal S / N}$\fi}
\def\siglos{\sigma_{\hbox{los}}}

\def\lae3{LAE~J095950.99+021219.1}

\title{
A Lyman-$\alpha$ Galaxy at Redshift $z=6.944$ in the COSMOS Field\thanks{This paper includes data gathered with the 6.5 meter Magellan Telescopes located at Las Campanas Observatory, Chile}
}

\author{
James E. Rhoads\altaffilmark{1},
Pascale Hibon\altaffilmark{1,2},
Sangeeta Malhotra\altaffilmark{1},
Michael Cooper\altaffilmark{3,4},
Benjamin Weiner\altaffilmark{5}
}

\begin{abstract}
  \lya\ emitting galaxies can be used to study cosmological
  reionization, because a neutral intergalactic medium 
  scatters \lya\ photons into diffuse halos whose
  surface brightness falls below typical survey detection limits.
  Here we
  present the \lya\ emitting galaxy \lae3, identified at redshift $z=6.944$ in
  the COSMOS field using narrowband imaging and
  followup spectroscopy with the IMACS instrument on the Magellan I
  Baade telescope.  With a single object spectroscopically confirmed
  so far, our survey remains consistent with a wide range of IGM
  neutral fraction at $z\approx 7$, but further observations are
  planned and will help clarify the situation.  Meantime, the object
  we present here is only the third  \lya-selected galaxy to be
  spectroscopically confirmed at $z \ga 7$, and is $\sim 2$--$3$ times
  fainter than the previously confirmed $z\approx 7$ \lya\ galaxies.
\end{abstract}

\altaffiltext{1}{School of Earth and Space Exploration,
Arizona State University, Tempe, AZ 85287; email malhotra@asu.edu,
James.Rhoads@asu.edu}
\altaffiltext{2}{Gemini Observatory, La Serena, Chile;
email phibon@gemini.edu}
\altaffiltext{3}{Center for Galaxy Evolution, Department of Physics
 and Astronomy, University of California, Irvine, 4129 Frederick
 Reines Hall, Irvine, CA 92697, USA; m.cooper@uci.edu}
\altaffiltext{4}{Hubble Fellow}
\altaffiltext{5}{Steward Observatory, University of Arizona}


\section{Introduction}

\lya\ emitting galaxies provide a valuable probe of reionization,
because resonant scattering of \lya\ photons
in the intergalactic medium (IGM) can suppress the
observed \lya\ line by a factor of $\ga 3$
for any neutral fraction $\ga 50\%$ \citep{mr04,Santos04}.
Such strong flux suppression will cause a
change in the \lya\ luminosity function that should be
obvious--- especially since \lya\ galaxies
show little evolution at $3\la z \la 6$, either in luminosity function 
\citep{Dawson07,Zheng12} or physical properties \citep{Malhotra12}.
Early studies concluded that
the IGM neutral fraction was already small at $z\approx 6.5$
\citep{Malhotra2004,Stern2005}.  More recent work has established the
\lya\ luminosity function at both $z\approx 5.7$ and $6.5$ to
considerable accuracy \citep{Ouchi10,Hu10,Kashikawa11},
showing a modest but statistically significant difference
between these observed \lya\ luminosity functions: 
The $z=6.5$ luminosity function is
below that at $z=5.7$, and the difference can be adequately characterized
by a pure luminosity evolution by a factor of $\sim 1.3$ \citep{Ouchi10}.

Yet, other \lya\ based tests for reionization---
including the apparent spatial clustering of \lya\ galaxies
\citep{McQuinn07}, the minimum ionized volume around observed \lya\
sources \citep{mr06}, and \lya\ line profiles \citep{Hu10,Ouchi10}---
show little evidence for neutral gas at $z\approx 6.5$.
This leaves an open question--- is lower
\lya\ luminosity function at $z=6.5$ due to neutral gas, or is it
an intrinsic evolution in the galaxy populations?

To distinguish between these possibilities, we can look to still
higher redshifts, where the IGM neutral fraction 
should be higher and its effects on \lya\ 
stronger.  The highest redshift readily accessible to \lya\ searches
using CCDs is $z\approx 7.0$, in the 9650\AA\ window in the
night sky OH emission spectrum.  We are pursuing a 
9650\AA\ narrowband survey using the IMACS imaging spectrograph on the 6.5m
Magellan~I Baade Telescope at Las Campanas Observatory \citep{Hibon11a}. 
We surveyed 465 square arcminutes, corresponding to 
$\sim 72000 \textrm{Mpc}^{3}$. After a
careful selection, we found 6 \zs\ LAE candidates \citep{Hibon11a}. 
To confirm whether these are real LAEs, we obtained 
multi-object spectra with IMACS.
In this {\it Letter}, we present the spectrum of \lae3, which 
was identified as a candidate redshift $z\approx 7$ \lya\ emitting
galaxy (candidate LAE 3) in \citet{Hibon11a}.   Our spectroscopy
reveals a single, isolated \lya\ line at redshift $z= 6.944$.  

Throughout the paper, we adopt a $\Lambda$CDM ``concordance
cosmology''  with $\Omega_M=0.27$, $\Omega_\Lambda=0.73$, and
$H_0=71 \kmsMpc$.


\section{Spectroscopic Observations and Analysis}
\subsection{Observations}
We observed our candidate $z\approx 7$ \lya\ galaxies using the
Inamori Magellan Areal Camera and Spectrograph (IMACS) on the 6.5m
Magellan~I Baade Telescope on the nights of 29--30 December 2010, and
8 February 2011.  The February data were of lower quality
and are not used here.
We used custom multi-slit masks, shared between two primary
observing programs.  We selected the f/2 camera and the 300-line
red-blazed grism with 1'' slitlets as the best compromise between 
areal coverage, spectral coverage, and spectral resolution.

Observations were split among five slit masks (two per night in
December, and one in February).  The time per mask and observing
conditions are summarized in table~\ref{mask_obs_table}.  While the
position angle of the masks were not all identical, the data were
taken without dithering the telescope.  Moreover, the targets were
centered on their slitlets, and \lae3\ is compact compared to the
seeing.  This allows us to combine all of the spectroscopic data into
a single 1D spectrum (see below).

\begin{table}
\begin{tabular}{lllll}
Mask ID & Observation & Number of & Time per & Seeing \\
        & date (UT)   & exposures & exposure & (approx) \\
\hline
COSMOS1 & 30 Dec 2010 & 4 & 1800 & 0.5--0.9$''$\\
COSMOS2 & 30 Dec 2010 & 3 & 1800 & 0.5--0.9$''$\\
COSMOS3 & 31 Dec 2010 & 4 & 1800 & 1--1.5$''$ \\
COSMOS4 & 31 Dec 2010 & 3 & 1860 & 1--1.5$''$ \\
COSMOS-Feb & 8 Feb 2011 & 5 & 1800 & 1.2--2$''$
\end{tabular}
\caption{Summary of slit mask observations.
 \label{mask_obs_table}}
\end{table}

\subsection{Data Reduction}
We performed initial data reduction steps using the COSMOS software
package\footnote{Note the double meaning of the acronym COSMOS.  We
  deny any responsibility for the ensuing confusion.}.
COSMOS steps include bias frame subtraction, spectroscopic flat
fielding using continuum (quartz) lamp exposures, and
wavelength calibration using arc lamp exposures. COSMOS also sky-subtracts
the spectra, using the \cite{Kelson03} algorithm
to remove night sky lines.  Finally, COSMOS extracts a 2D, rectified 
spectrum for each slitlet.

We performed subsequent steps in two ways, either (a) combining
exposures from each mask separately, and then combining results from
different masks; or (b) directly combining all exposures from multiple 
masks.

Treating masks separately gives four 2D spectra
of \lae3\ from December, and one more from February.
Most of these 2D stacks show a weak but visible emission line in the 
spectrum of \lae3.  We next combined the four December spectra  
into a stacked 2D spectrum comprising our best 7.05 hours of data.  
To do this, we first averaged the four 2D spectra. 
Next, we made a median-combined stack.  
We then subtracted the two, and computed the (sigma-clipped)
noise level in the difference.  Finally, we constructed a hybrid stack,
using the value from the average stack almost everywhere, but the 
value from the median stack wherever the difference between 
these two stacks exceeded $10\sigma$. This yields a lower noise 
estimate than the median, yet remains
more robust to outliers than the mean.
The emission line becomes readily evident in this combined stack.

\begin{figure}[h]
\plotone{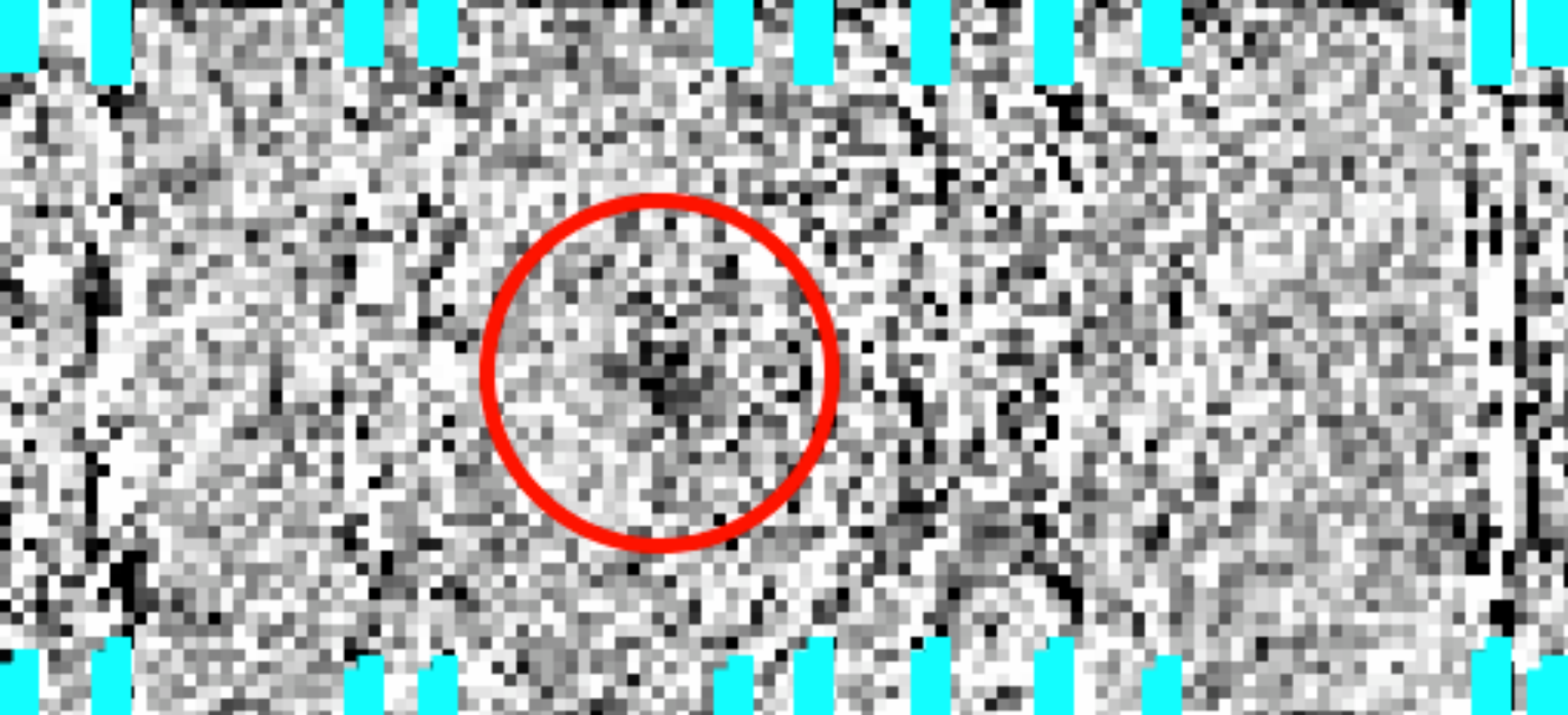}
\plottwo{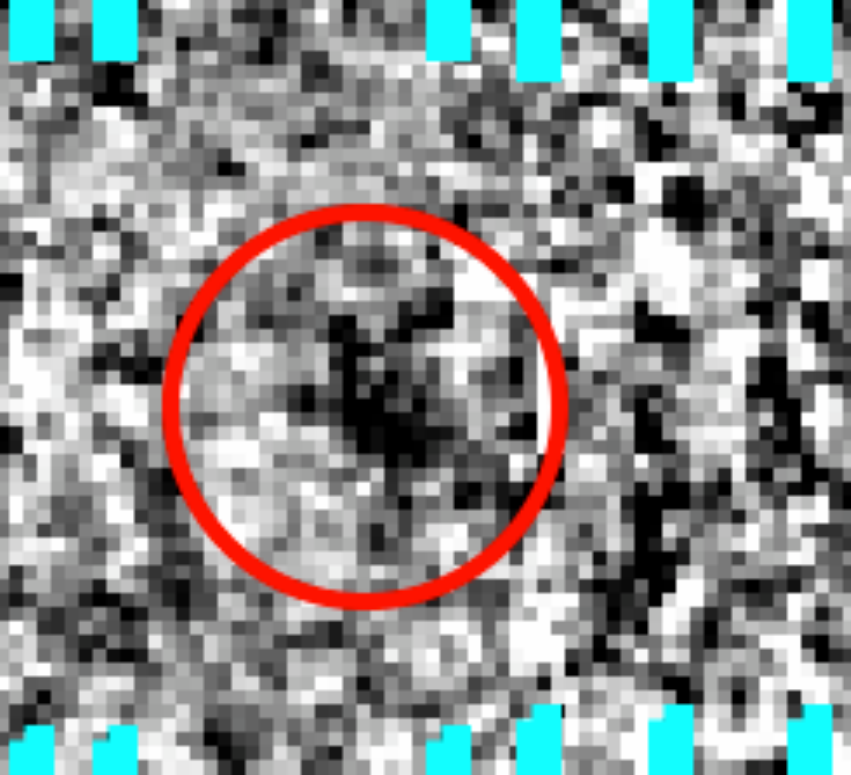}{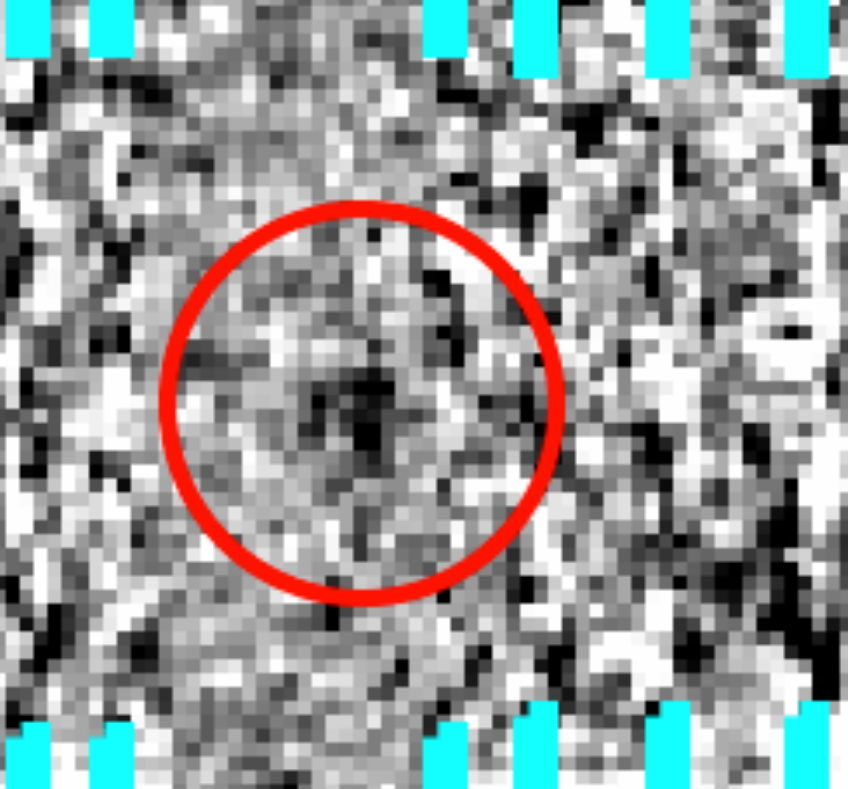}
\caption{2D spectra of \lae3.  {\it Upper panel:\/} Full December data
  set (14 exposures, 7.05 hours' integration) stacked together.  
  Wavelength increases from 9550\AA\ at
  left to 9800\AA\ at right.  The spatial extent is
  $12''$ from bottom to top. 
  {\it Lower panels:\/} First night (left) and second night (right) of data
  separately.  The lower 
  panels have been lightly smoothed for clarity, with a Gaussian kernel
  having $\sigma=0.6$ pixel.
  Cyan bars at top and bottom mark the wavelengths of night sky emission
  lines, with longer bars denoting brighter lines.
  \label{fig:2dspec}}
\end{figure}

To test the robustness of our results, we also combined all December
exposures in single 14-frame stacks, using various outlier rejection
schemes (median stacking with 3- and 5$\sigma$ rejection, and
average stacking with 2.5$\sigma$ rejection). The emission line remains
comparably significant in all of these stacks.  The stacked
2D spectrum around the emission line, using average
stacking and $2.5\sigma$ rejection, is shown in figure~\ref{fig:2dspec}.

We also made stacks by bootstrap resampling, stacking
14 exposures selected randomly with repetitions permitted.
We remeasured the flux at the location of the detected
line, using aperture photometry in the 2D stacks.  The bootstrap
fluxes were $104\% \pm 10\%$ of the ``normal'' stack flux 
for a 10-pixel diameter ($2''\times 20$\AA) aperture,
and $116\% \pm 21\%$ for a 14-pixel diameter ($2.8'' \times 28$\AA)
aperture.  Among 1000 bootstrap simulations, the lowest measured 
fluxes were 77\% and 66\% of ``normal'' for the 10- and 14-pixel apertures. 
The observed line is therefore not a fluke caused by a handful of 
exposures.

\begin{figure}[h]
\plotone{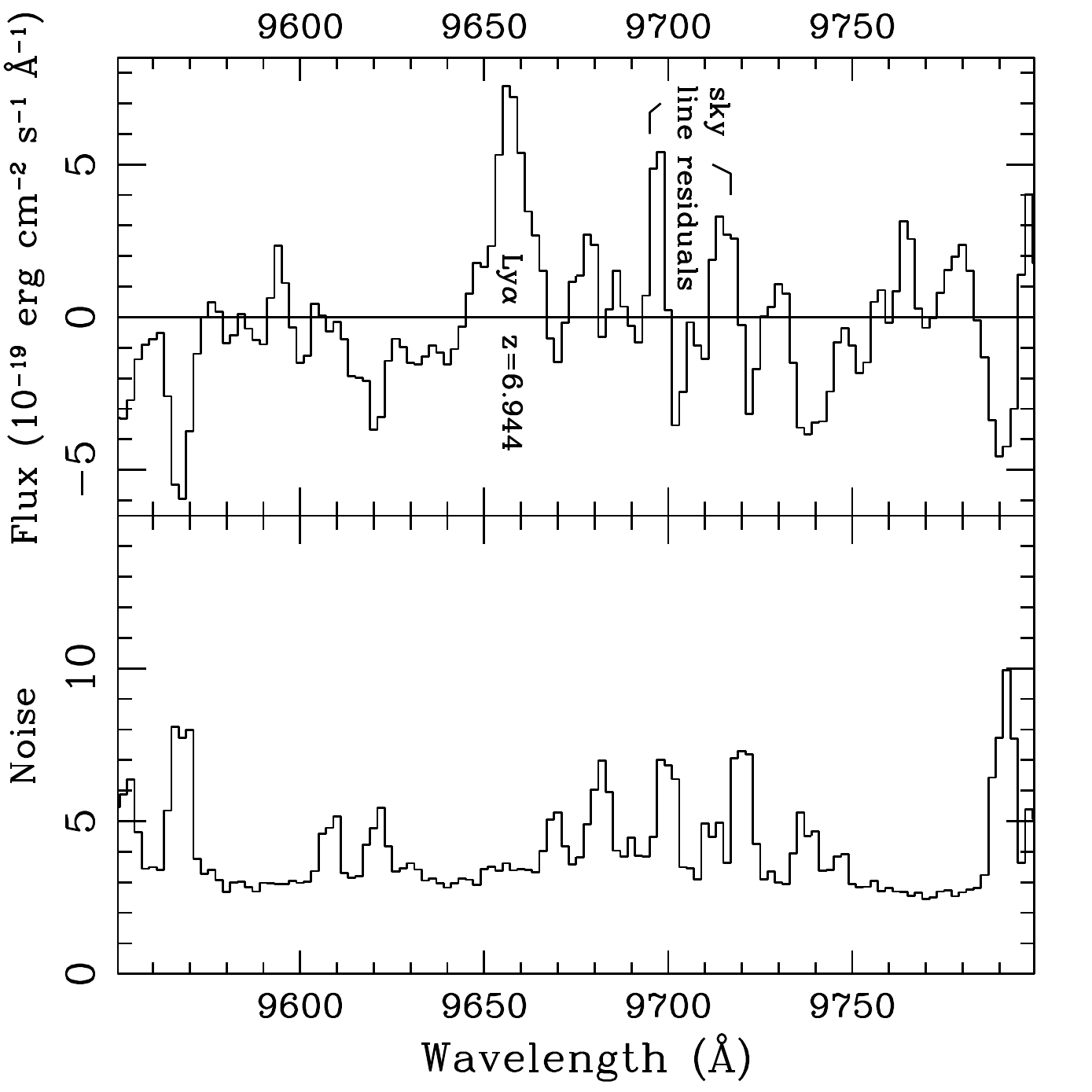}
\caption{
The extracted 
1D spectrum (lightly smoothed with a ``1 2 1'' filter, with 4\AA\ FWHM)
at top, and a noise estimate (unsmoothed) at bottom.  The sole
significant feature is the \lya\ line at 9657\AA,
corresponding to redshift $z=6.944$.  Some residuals from subtraction
of the strong night sky OH emission lines remain, and are marked.
The most notable of these is the spike at 9698\AA.  The plotted noise
is based on the sigma image from the 14-frame stack,
scaled for the number of spatial pixels combined in the 
1D extraction.
  \label{fig:1dspec}
}
\end{figure}

We next extracted a 1D spectrum from the two-stage stacking (method ``a''), 
using the IRAF task ``apall''  with an unweighted extraction of $1.4''$
(7 pixel) window width, centered on the emission line
and parallel to the dispersion axis.  (COSMOS 2D output
has the dispersion axis parallel to the $x$-axis, so we need not 
fit a trace to the continuum, which is undetected in
the present data anyway.)  We performed no further sky subtraction,
since that too is done by the COSMOS package.  To get another estimate
of the noise level in the data, we extracted five further 1D spectra
from the 2D stacked spectrum, each at a different
spatial position along the slit.  Each should be essentially a pure
noise spectrum.  The variance among these five parallel traces provides
a wavelength-dependent noise estimate, and the line is significant
at the $4.5\sigma$ level against this estimate.
The extracted 1D spectrum is shown in figure~\ref{fig:1dspec}.

\paragraph{Significance:}
To explore the significance of the line detection, we measured
aperture fluxes at a grid of clean locations in the 2D spectrum of
\lae3\ (after rescaling the 2D spectrum by the noise ratio
$\sigma(9658\AA) / \sigma(\lambda)$).  The RMS counts among these
apertures corresponds to a $1\sigma$ noise of $1.34\times
10^{-18}\ergcm2s$, against which our line is a $6.3\sigma$ event.
Among $\ga 400$ non-overlapping 5-pixel apertures, the brightest two
were $4.1\sigma$ and $3.2\sigma$ events (65\% and 51\% as bright as the
\lae3\ line), suggesting only mildly non-gaussian noise.

The search that found this line was based on 6 candidates, each with a
position known to $1''$ and an expected line wavelength known to
$\Delta\lambda=90$\AA.  Given our spatial and spectral resolution,
this corresponds to $\sim 6\times 2\times 20=240$ independent
resolution elements.  Our significance level estimates range from
$4.5\sigma$ to $6.3\sigma$, and for Gaussian noise, the corresponding
chance probabilities in 240 trials range from $<10^{-3}$ to
$<10^{-7}$.

\paragraph{Spectroscopic flux calibration:}
Each mask included two blue stars with well
measured photometry from the COSMOS project \citep{Capak07}. 
We flux calibrate the observed emission line of \lae3\ by direct
comparison with the observed counts in one of these stars, 
which was observed under identical conditions as our science targets.
Both the emission line count rate and the comparison star's count rate
per unit wavelength were measured directly in the 2D spectra and at
the same wavelength.  For the emission line, we used the 10-pixel
diameter ($2''\times 20$\AA) aperture. This yields a flux of
$8.5\times 10^{-18} \ergcm2s$ for \lae3, corresponding to line
luminosity $4.9\times 10^{42} \ergsec$ at $z=6.944$.
  
The fractional uncertainties in this flux are $\sim 20\%$
from photon counting statistics (with a statistical signal-to-noise ratio
of $s/n=4.5$--$6.3$ from the 1D or 2D spectra), $12\%$ from the choice of
aperture used to measure the line flux in the 2D spectrum, and 
$\le 10\%$ from the assumption that the comparison star's
flux density at 9680\AA\ equals its \textit{z}-band flux density.  
The final spectroscopic flux measurement is
$(8.5 \pm 2) \times 10^{-18} \ergcm2s$.

\subsection{Comparison to narrowband imaging results:}
The observed spectroscopic line flux is smaller than the
narrowband flux \citep{Hibon11a}.  Part of the
discrepancy can be attributed to continuum in
the narrowband filter.  The emission line is near the blue
edge of the filter bandpass, so continuum emission redward of the
line will contribute relatively strongly to the narrow band flux.

We also re-examined the narrowband flux measurements for \lae3. 
Following \citet{Hibon11a}, we used moderately bright
stars that are well detected but unsaturated in both the public
COSMOS \textit{z}-band image and our NB9680 image.  The narrowband
magnitudes in \citet{Hibon11a} were based on $1''$ diameter aperture
fluxes, with an aperture correction based
on the difference between SExtractor ``magiso'' and aperture fluxes
(see \citet{Hibon11a} for more details).  
In the present work, we
omitted the aperture correction step, using instead identical 
$1.2''$ diameter flux measurements for both the science objects and the
reference stars in the NB9680 image.  (The precise aperture diameter
is unimportant, since \lae3\ is compact compared to the point
spread function.)  We obtained
a narrowband magnitude $NB9680_{AB}=24.86\pm 0.18$.  The corresponding
narrowband flux is $12\times 10^{-18}\ergcm2s$.

This is almost 50\% more than the spectroscopically
determined emission line flux (a $1.5\sigma$ difference). 
If we attribute the difference to
continuum emission in the narrowband filter, the corresponding
flux density is $f_\lambda\sim (f_{NB}-f_{spec})/\Delta\lambda
\approx 3.5\times 10^{-18} / 60\hbox{\AA}\approx 5.8\times 10^{-20}\flam_unit$,
or $AB=25.75$. 
The implied observer-frame equivalent width is 
$\sim 120$\AA\ (though consistent with an arbitrarily large
equivalent width at $1.5\sigma$).  While the object
is undetected in the $z'$ filter down to $AB > 26.4$,
a continuum magnitude of $25.75$ {\it redward of the line at 9657\AA\ } 
remains allowed, since most of the z$^\prime$ filter's
transmission lies blueward of that wavelength.  The object is also undetected 
in the WIRDS J-band image \citep{Bielby12}, 
with a $1.2''$ aperture flux 
$\sim(1.8\pm 2.1)\times 10^{-30}\fnu_unit$, corresponding to 
$J_{AB}\ga 24.1$ ($3\sigma$).

\section{Interpretation}
We interpret the line at 9657\AA\ in \lae3\ as \lya\ at redshift $z=6.944$,
based on non-detections in all filters blueward of this line, and 
on the absence of other optical lines.

\begin{figure}[h]
\plottwo{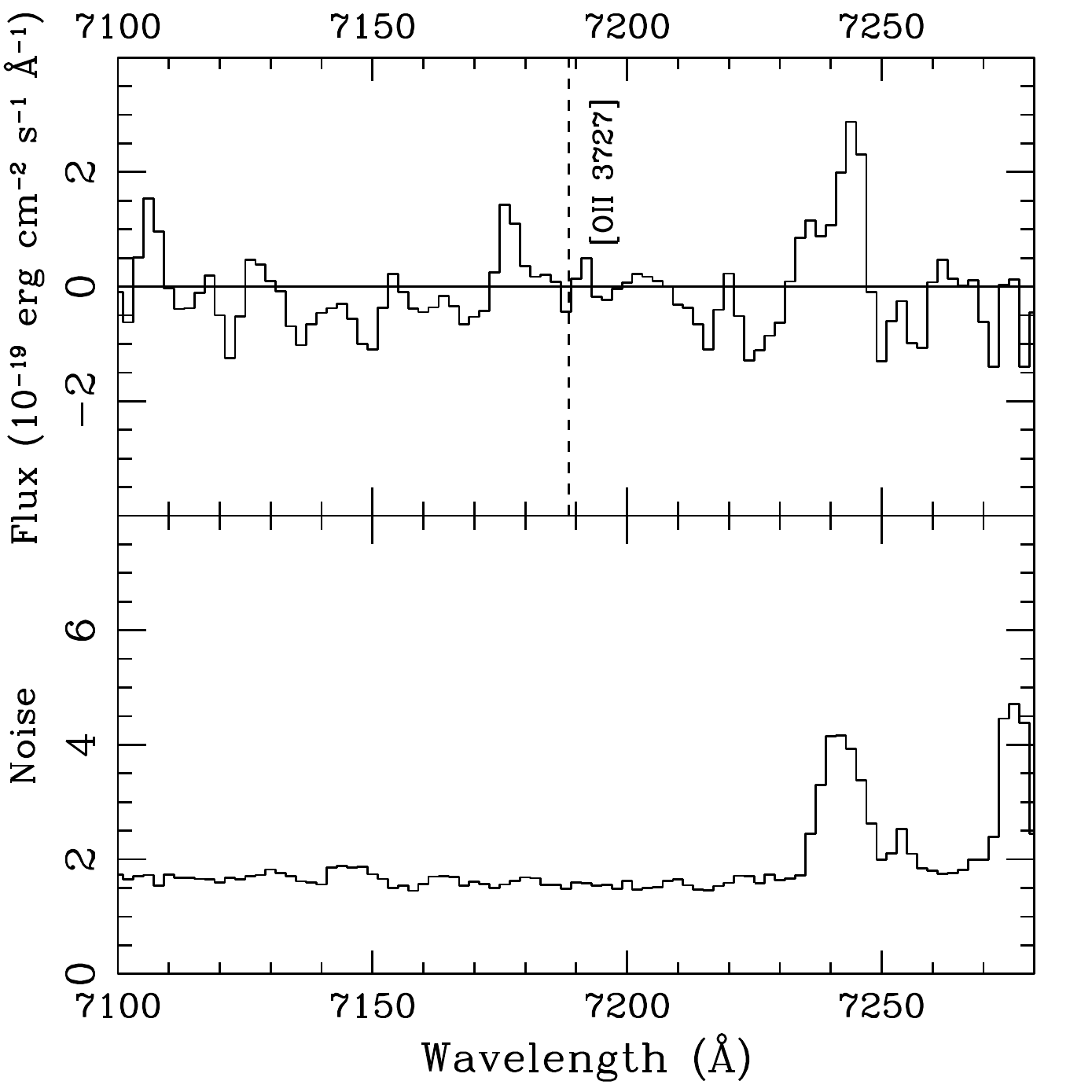}{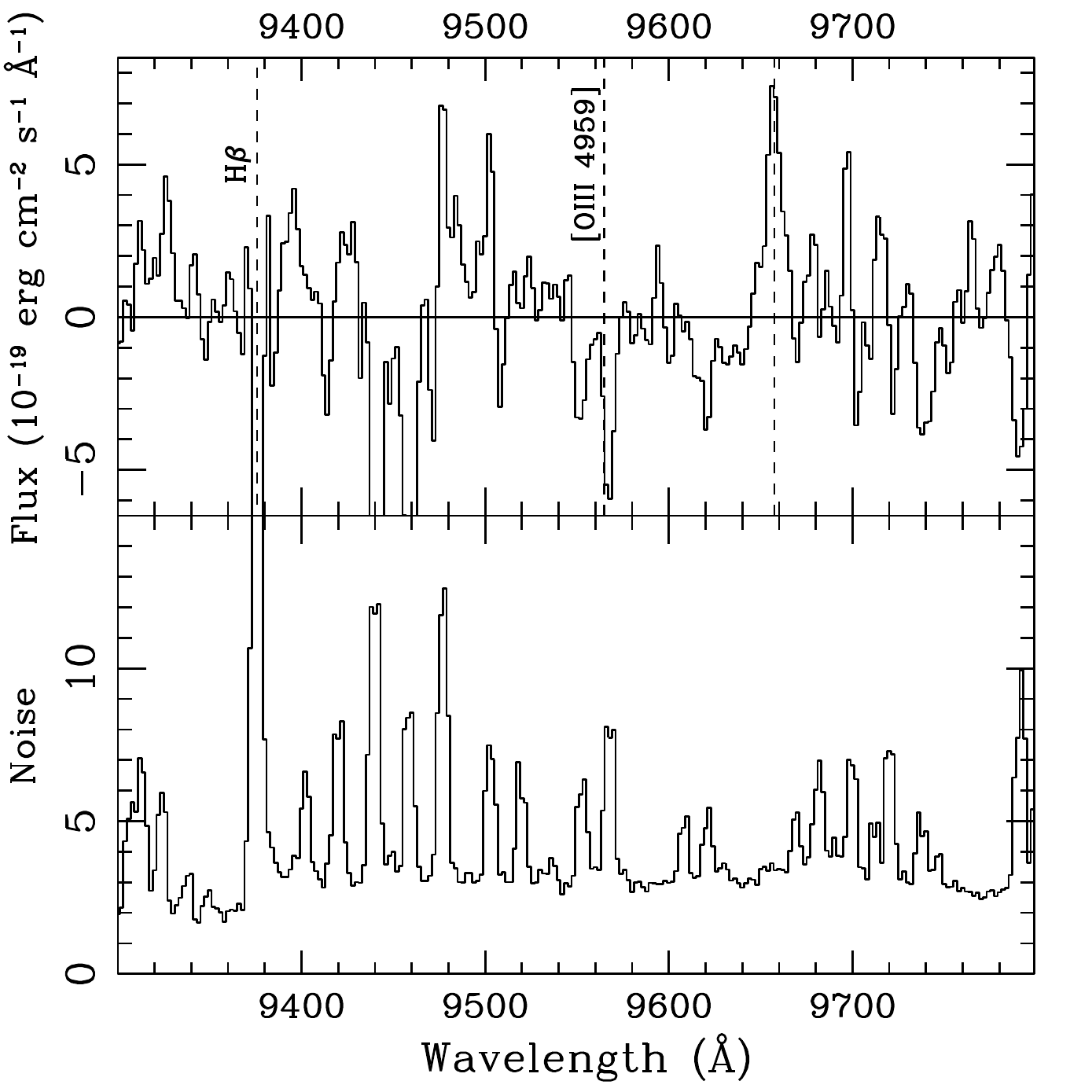}
\plottwo{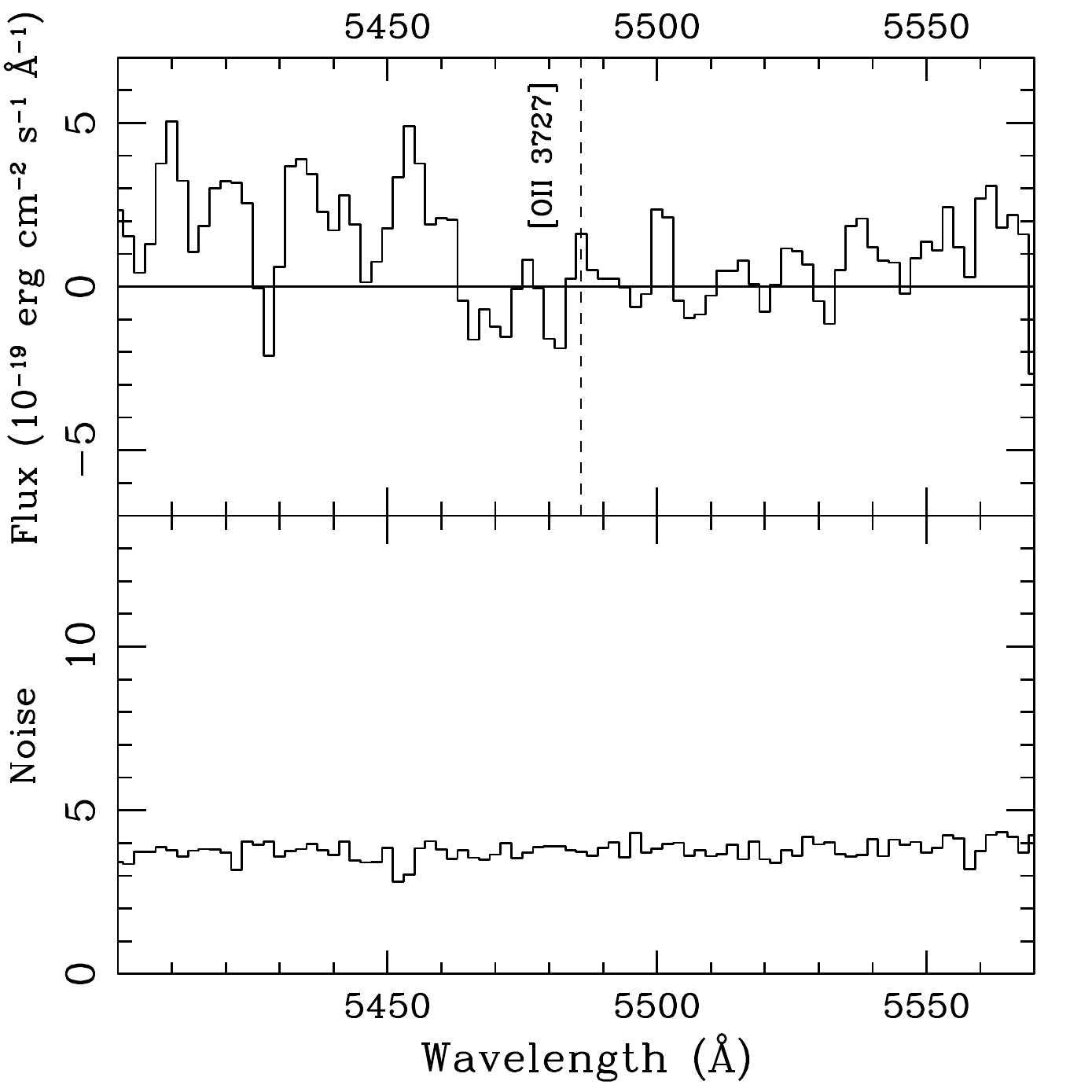}{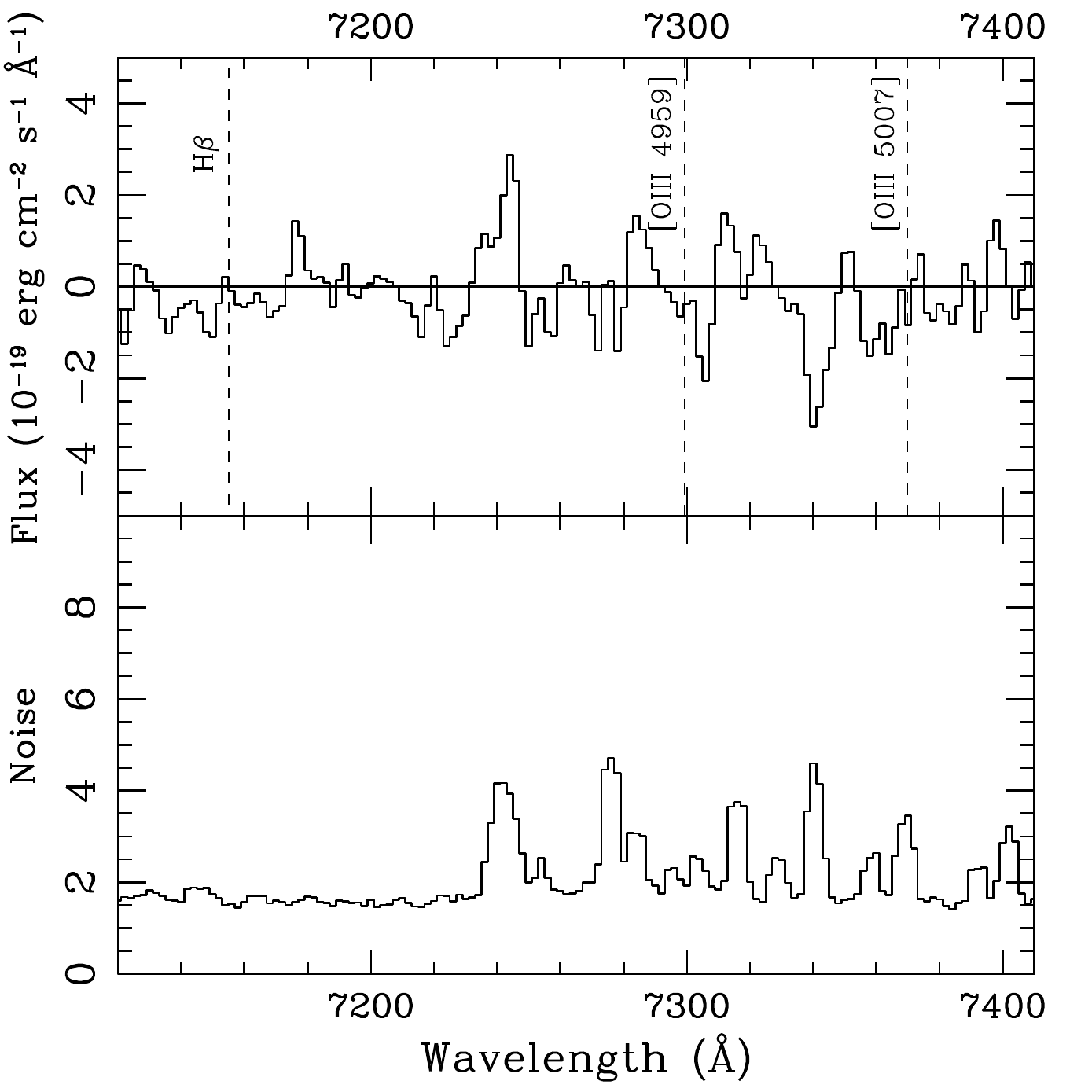}
\caption{If the emission line in \lae3\ were not \lya\ but instead
were H$\alpha$ or \oiii, we should expect additional emission lines 
in our optical spectrum.  These figures show the expected locations of 
the \oiiipair, H$\beta$, and \oii\ lines, for the cases where the primary 
line is either \oiii\ at $z = 0.929$ (top two panels) or H$\alpha$ 
at $z= 0.472$ (bottom two panels).  
In no case do we see a prominent emission line, although 
for the $z=0.929$ case the \oiiib\ and H$\beta$ lines both
fall on night sky line residuals, leaving the absence of an \oii\ line 
at 7188\AA\ and the lack of any optical broad band detection as the 
best discriminants against this possibility.  As in fig.~2, the 
signal spectra are lightly smoothed with a (1 2 1) filter, while the noise
spectra are unsmoothed.
\label{fig:fg_check}
}
\end{figure}

Were the primary line H$\alpha$ (at $z\approx 0.472$), 
or \oiii\ (at $z\approx 0.928$), we would expect other
prominent emission lines in our IMACS spectrum.
Figure~\ref{fig:fg_check} shows non-detections in the 1D spectrum at the
expected line wavelengths. 
Corresponding upper limits are summarized in table~\ref{tab:fglinelimit}. 
The ``H$\alpha$'' case ($z\approx 0.472$)
is disfavored by the non-detections of \oiii\ and \oii, with $3\sigma$  
line ratio
limits $f($\ion{O}{3}$)/f({\rm H}\alpha) < 1/2$ and 
$f($\ion{O}{2}$)/f({\rm H}\alpha) < 2/3$.
If the primary line is \oiii,
unfortunately placed night-sky line residuals overlap the 
expected locations of \oiiib\ and H$\beta$, precluding interesting limits.
Fortunately, the expected \oii\ line location is clean, and gives
a tight upper limit $f($\ion{O}{2}$)/f($\ion{O}{3}$) < 0.2$ ($3\sigma$).
This provides some evidence against the \oiii\ interpretation. 
Ratios of $f($\ion{O}{2}$)/f($\ion{O}{3}$)$ 
this small are seen in a significant
minority of star forming galaxies \citep{Xu07,McLinden11,Richardson11,Xia11},
but more can be ruled out by this line ratio.

\begin{table}
\begin{tabular}{lllrc}
If $z=$... & Secondary Line & Expected & Formal flux & $3\sigma$ limit\\
& & wavelength & \multicolumn{2}{c}{ ($10^{-18} \ergcm2s$) } \\
0.472 & \oiii\ & 7370\AA\ & $-0.4\pm 1.5$ & 4.1 \\
0.472 & \oiiib\ & 7299\AA\ & $-0.6\pm 0.8$ & 1.8 \\
0.472 & H$\beta$ & 7154\AA\ & $-0.2 \pm 0.8$ & 2.2 \\
0.472 & \oii\  & 5486\AA\ & $0.6 \pm 1.6$ & 5.4 \\
0.928 & \oiiib\ & 9565\AA\ & $-4.3 \pm 4.6$ & 9.5 \\
0.928 & H$\beta$ & 9375\AA\ & $-12 \pm 15$ & 33 \\
0.928 & \oii\ & 7188\AA\ & $0.0 \pm 0.5$ & 1.5 
\end{tabular}
\caption{Line flux limits at locations where we would expect emission lines, 
were the primary line at 9657\AA\ not \lya.   $z=0.472$ corresponds 
to H$\alpha$ at 9657\AA, and $z=0.928$ corresponds to 
\oiii\ at 9657\AA. 
For comparison, the 9657\AA\ line has flux $8.5\times 10^{-18}\ergcm2s$. 
Limits tabulated here are based on fitting a Gaussian line profile to the
observed spectrum, with the central wavelength and line width for the fit
fixed by the properties of observed line.
\label{tab:fglinelimit}}
\end{table}

To address the \oii\ possibility and 
further improve our constraints on \oiii, we examine equivalent
widths.  Following \citet{Hibon11a}, we combine our spectroscopic line flux
with 
optical magnitude limits of 27.9, 27.6, and 27.3 mag
($5\sigma$, AB) in g$^\prime$, r$^\prime$, and i$^\prime$ filters
respectively.  Since star-forming galaxies have $f_\nu \sim 
\hbox{constant}$, we have
$f_\lambda(9657{\rm \AA}) \approx 3.6\times 10^{-20} \times 
10^{-0.4\times 28}\times c/(9657{\rm \AA})^2=7.3\times 10^{-21}\flam_unit$.
The nondetection in these optical images then implies a
$5\sigma$ limit 
$EW\equiv f_{_{\rm line}}/f_\lambda 
\approx (8\times 10^{-18}/7.3\times 10^{-21}) {\rm \AA} = 1100{\rm \AA}$
(observer frame).

While \oiii\ and H$\alpha$ emission line sources with 
equivalent widths this large exist
\citep{Rhoads00,Kakazu07,Straughn08,Straughn09,vdWel11,Atek11},
they are exceptional, rare objects.  In \citet{Hibon11a},
we estimated the numbers expected in our survey based on 
published line luminosity functions \citep{Kakazu07,Geach10}
and equivalent width distributions \citep{Straughn09}.
We found that $\la 0.3$ \oiii\ emitters and $\la 0.6$ H$\alpha$ emitters
are expected.

\oii\ emitters have generally smaller equivalent widths. 
We found {\it no} \oii-selected objects with  
$EW \ga 1100 {\rm \AA} / (1+z_{[OII]}) \approx 425 {\rm \AA}$ in
the samples from \citet{Straughn09} (30 objects), \citet{Kakazu07} (24
objects), \citet{Xia11} (11 objects), or \citet{Drozdovsky05} (400
objects).  Thus, $\la 1/465 = 0.0022$ of 
\oii\ emitters might enter our sample as LAE candidates.  The luminosity
function from \citet{Rigopoulou05} suggests that our survey
volume should contain $\sim 45$ \oii\ galaxies.  
Among these, $\la 0.1$ object should pass our \lya\ selection criteria.

Thus, the aggregate sample of foreground emitters expected in our
survey is $< 1$ galaxy.  In contrast, our survey volume at $z\approx
6.95$ should contain between $\sim 2.5$ and $\sim 11$ \lya\
galaxies with line fluxes $\ga 8\times 10^{-18} \ergcm2s$, based on
the $z\approx 6.5$ luminosity functions of \citet{Hu10} and
\citet{Ouchi10}.  We thus regard \lya\ as the best interpretation
of the observed emission line.


\section{Discussion}
Redshift $z\approx 7$ is the current frontier in reionization studies, 
an area of active exploration where our observational knowledge is
growing rapidly.

The recently discovered quasar at $z\approx 7.1$ (Mortlock et al
2011), combined with spectroscopic followup (and occasional
confirmation) of $z\sim 7$ galaxy candidates from HST
WFC3 surveys, provide an unprecedented look at this epoch.
Observations of these objects seem to favor the continued existence of
significant neutral intergalactic gas as late as $z\sim 7$.
This is surprising, given that microwave background polarization data
from WMAP favor a characteristic reionization redshift $\zre \sim 11$, 
and that the IGM at $z\sim 6.2$ is highly ionized, with a neutral fraction 
of only 1--4\% based on quasar spectra (Fan et al 2006).  Nonetheless, the
ionized bubble around the $z\approx 7.1$ quasar appears too small to
be comfortably explained in a fully ionized medium, unless the quasar
is itself remarkably young ($\sim 10^6$ years) (Bolton et al 2011).
Similarly, three independent research groups have argued that the
fraction of $z\sim 7$ Lyman break candidates showing \lya\ emission
appears smaller than would be expected in an ionized medium (Vanzella
et al 2010; Stark et al 2010; Ono et al 2011; Schenker et al 2011;
Pentericci et al 2011).  Still, these results depend on the reliability
of photometric selection criteria, and a contamination of order 50\%
could explain the observations without recourse to neutral 
gas (e.g., Schenker et al 2011).

\lya\ galaxy surveys offer a complementary approach to studying reionization.
The underlying physics is the same as for spectroscopic followup of 
Lyman break samples, but the survey selection proceeds differently, 
leading to different potential selection biases.  The uncertainties in the
method are likewise very different from those associated with the
quasar near-zone measurement \citep{Mortlock11,Bolton11}, or 
the Gunn-Peterson trough \citep{Fan06}.  Because of this,
conclusions about cosmological reionization will be strongest when
they are based on multiple independent methods.   

The work we present here is only the second large-area narrowband
survey for for \lya\ galaxies at redshift $z\approx 7.0$, following on
the work of \citet{Iye06} and \citet{Ota10}.  The 
spectroscopic confirmation of \lae3\ demonstrates that such
objects can be identified at flux levels considerably fainter than the
$2\times 10^{-17} \ergcm2s$ line of IOK-1 \citep{Iye06}
or the recently reported $z\approx 7.2$ narrowband-selected 
\lya\ galaxy SXDF-NB1006-2 \citep{Shibuya11}.  Our observed
line flux, $8.5\times 10^{-18} \ergcm2s$, corresponds to a rest-frame
line luminosity of only $\llya = (4.9 \pm 1) \times 10^{42} \ergsec$.
This is at or below the characteristic line luminosity
$L^*_{Ly\alpha}$ from Schechter function fits to $z\approx 6.5$ \lya\
samples (e.g., $L^*_{Ly\alpha} = (4.4 \pm 0.6) \times 10^{42} \ergsec$
and $\Phi^* = 8.5 \times 10^{-4} \cMpc^{-3}$, \citet{Ouchi10}; 
$L^*_{Ly\alpha} = 1.0\times 10^{43} \ergsec$ and 
$\Phi^* =6\times 10^{-5} \cMpc^{-3}$,\citet{Hu10}).

Thus, our survey has achieved sensitivity to typical \lya\ emitters.
Were the $z=7$ luminosity function unchanged from that at $z=6.5$, we
would expect our survey volume to contain 11 or 2.6 \lya\ emitters
brighter than \lae3, based on the LF of \citet{Ouchi10} or
\citet{Hu10} respectively. 
This reflects large differences in expectations from different published
$z=6.5$ luminosity functions.
We will present a detailed
analysis of our survey's constraints on the \lya\ luminosity function
in a future paper, after we have more sensitive spectra for our
remaining candidates.  For now, the galaxy \lae3\ is among the few
most distant spectroscopically confirmed galaxies known, and the
faintest to be discovered through a direct search for \lya\ line
emission.

\acknowledgments JER, PH, and SM gratefully acknowledge support of 
the National Science Foundation through NSF grant AST-0808165.
MCC acknowledges support from 
NASA through Hubble Fellowship grant \#HF-51269.01-A, awarded by the
Space Telescope Science Institute, which is operated by the
Association of Universities for Research in Astronomy, Inc., for NASA,
under contract NAS 5-26555; and 
from the Southern California Center for Galaxy Evolution, a
multi-campus research program funded by the University of California
Office of Research. 
We thank the staff of Las Campanas Observatory for their expert 
assistance throughout this project, and Gabriel Martin and
David Osip in particular for their help with our filters. 
Finally, we thank the referee for a thorough and fair report.


\end{document}